\documentclass[aps,pra,twocolumn,showpacs,superscriptaddress]{revtex4-1}
\usepackage{graphicx}
\usepackage{amsmath,amsfonts}
\usepackage[usenames]{color}

\newcommand{\tauQ}{\tau_{\rm Q}}

\def\be{\begin{equation}}
\def\ee{\end{equation}}
\def\bea{\begin{eqnarray}}
\def\eea{\end{eqnarray}}
\def\bi{\begin{itemize}}
\def\ei{\end{itemize}}

\begin{document}

\title{Dynamics of the modified Kibble-\.Zurek mechanism in antiferromagnetic spin-1 condensates}

\author{Emilia Witkowska}
\affiliation{Instytut Fizyki PAN, Aleja Lotnik\'ow 32/46, 02-668 Warsaw, Poland}

\author{Jacek Dziarmaga}
\affiliation{Instytut Fizyki Uniwersytetu Jagiello\'nskiego, ul. Reymonta 4, 30-059 Krak\'ow, Poland}

\author{Tomasz \'Swis\l{}ocki}
\affiliation{Instytut Fizyki PAN, Aleja Lotnik\'ow 32/46, 02-668 Warsaw, Poland}

\author{Micha\l{} Matuszewski}
\affiliation{Instytut Fizyki PAN, Aleja Lotnik\'ow 32/46, 02-668 Warsaw, Poland}

\begin{abstract}
We investigate the dynamics and outcome of a quantum phase transition from an antiferromagnetic 
to phase separated ground state in a spin-1 Bose-Einstein condensate of ultracold atoms. 
We explicitly demonstrate double universality in dynamics within experiments with various quench time.
Furthermore, we show that spin domains created in the nonequilibrium transition constitute a 
set of mutually incoherent quasicondensates. The quasicondensates appear to be positioned in a semi-regular fashion, 
which is a result of the conservation of local magnetization during the post-selection dynamics. 
\end{abstract}

\maketitle

\section{Introduction}

One of the great achievements of statistical mechanics is the ability to describe complex systems of many
particles using a limited set of variables describing collective behavior. 
Consequently, the complicated microscopic dynamics of the system
is reduced to tractable models. Universality of phase transitions is a particularly striking example of such reduction,
where the multitude of physical models are divided into a finite number of 
universality classes characterized by certain symmetry properties and critical scaling laws.
While theoretical description of universality of equilibrium phase transitions is provided by the renormalization 
group~\cite{RenormalizationGroup}, universality in nonequilibrium systems is not yet fully understood~\cite{NonequilibriumBooks}.

A system that is normally in an equilibrium state may become out of equilibrium when it is driven through a 
second order phase transition, due to the divergence of the relaxation time. 
If symmetry breaking occurs at the same time, the transition
may result in the creation of defects, such as domain walls, vortices or strings.
This process, called the Kibble-\.Zurek mechanism (KZM), was predicted in a number of physical systems,
including the dynamics of the early Universe~\cite{Kibble,Zurek}, 
and observed in experiments with superfluid Helium~\cite{Helium_KZ}, liquid crystals~\cite{other_KZ},
superconductors~\cite{superconductors_KZ}, cold atomic gases~\cite{BEC_KZ},
Dicke quantum phase transition~\cite{Esslinger}, and most recently in ion traps \cite{Ulms}. Importantly, the Kibble-\.Zurek 
theory predicts the universality of dynamics of nonequilibrium phase transitions.

Quantum phase transition, in contrast to a classical (thermodynamic) one,
occurs when varying a physical parameter leads to a change of the nature of the ground state~\cite{QPT_Book}.
Recently, a few theoretical works demonstrated that the KZM can be successfully applied to describe quantum phase
transitions in several models~\cite{Other_QPT,Spinor_FerromagneticKZ}, 
see Ref. \cite{Nonequilibrium} for reviews. 
Among these, Bose-Einstein condensates of ultracold atoms
offer realistic models of highly controllable and tunable systems~\cite{Spinor_FerromagneticKZ}. 

In a recent paper~\cite{Nasz_PRL}, we demonstrated that the quantum phase transition from an antiferromagnetic 
to phase separated ground state in a spin-1 Bose-Einstein condensate of ultracold atoms exhibits scaling laws
characteristic for systems displaying universal behavior on various length scales. Phase separation
leads to the formation of spin domains, with the number of domains dependent on the quench time.
Interestingly,
the Kibble-\.Zurek scaling law was obtained only for the dynamics close to the critical point. Further on,
the post-selection of domains was observed, which gave rise to a second scaling law with a different exponent.
The post-selection was attributed to the conservation of an additional quantity, namely the condensate magnetization.

In this paper, we describe in detail the dynamics of this phase transition. 
For simplicity, we consider a system in the ring-shaped 1D geometry with periodic boundary conditions.
By employing the Bogoliubov approximation
in both the initial and the phase separated state, we derive the scaling laws observed numerically and explain
the post-selection process. We explicitly demonstrate universality in dynamics within experiments with various quench 
times by employing appropriate scalings of space and time.
Furthermore, we show that spin domains created in the nonequilibrium transition constitute a 
set of mutually incoherent quasicondensates. The quasicondensates appear to be positioned in a semi-regular fashion, 
which is a result of the conservation of local magnetization during the post-selection dynamics. 

\section{The model and its phase diagram}

We consider a dilute antiferromagnetic spin-1 BEC in a homogeneous magnetic field pointing along the $z$ axis.
We start with the Hamiltonian $H = H_0 + H_{\rm A}$, where the symmetric (spin-independent) part is 
\begin{equation} \label{En}
H_0 = \sum_{j=-,0,+} \int d x \, \psi_j^\dagger \left(-\frac{\hbar^2}{2m}\nabla^{2} + \frac{c_0}{2} \rho 
+ V(x)\right) \psi_j. 
\end{equation}
Here the subscripts $j=-,0,+$ denote sublevels with magnetic quantum numbers along the magnetic field axis $m_f=-1,0,+1$,
$m$ is the atomic mass, $\rho=\sum \rho_j = \sum \psi_j^\dagger \psi_j$ is the total atom density, 
$V(x)$ is the external potential. 
Here we restricted the model to one dimension, with the other degrees of freedom confined
by a strong transverse potential with frequency $\omega_\perp$. The spin-dependent part can be written as
\begin{equation} \label{EA}
H_{\rm A} = \int d x \, \left[ \sum_j E_j \rho_j + \frac{c_2}{2} :{\bf F}^2:\right]\,,
\end{equation}
where $E_j$ are the Zeeman energy levels, the spin density is 
${\bf F}=(\psi^{\dagger}F_x\psi,\psi^{\dagger}F_y\psi,\psi^{\dagger}F_z\psi)$,
where $F_{x,y,z}$ are the spin-1 matrices and $\psi =(\psi_+,\psi_0,\psi_-)$.
The spin-independent and spin-dependent interaction coefficients are given by 
$c_0=2 \hbar \omega_\perp (2 a_2 + a_0)/3>0$ and $c_2= 2 \hbar \omega_\perp (a_2 - a_0)/3>0$, 
where $a_S$ is the s-wave scattering length for colliding atoms with total spin $S$.
In the following analytic calculations we often assume the incompressible regime where
\be 
c_0~\gg~c_2~,
\ee
which is a good approximation in the case of a $^{23}$Na spin-1 condensate, where $c_0\approx 32 c_2$.

The total number of atoms $N = \int \rho d x$ and magnetization $M = \int \left(\rho_+ - \rho_-\right) d x$ are conserved quantities. 
In reality, there are processes that can change both $N$ and $M$,
but they are relatively weak in $^{23}$Na condensates~\cite{Chang_NP_2005} and can be neglected on the time scales considered below.

The linear part of the Zeeman shifts $E_j$ induces a homogeneous rotation of the spin vector around the direction of the magnetic field.
Since the Hamiltonian is invariant with respect to such spin rotations, 
we consider only the effects of the quadratic Zeeman shift~\cite{Matuszewski_AF,Matuszewski_PS}.
For sufficiently weak magnetic field we can approximate it by a positive energy shift of the $m_f=\pm 1$ sublevels 
$\delta=(E_+ + E_- - 2E_0)/2 \approx B^2 A$, 
where $B$ is the magnetic field strength and $A=(g_I + g_J)^2 \mu_B^2/16 E_{\rm HFS}$,
$g_I$ and $g_J$ are the gyromagnetic ratios of electron and nucleus, $\mu_B$ is the Bohr magneton, 
$E_{\rm HFS}$ is the hyperfine energy splitting at zero magnetic field \cite{Matuszewski_AF,Matuszewski_PS}.
Finally, the spin-dependent Hamiltonian (\ref{EA}) becomes
\begin{equation} 
H_{\rm A} = \int d x \, \left[ AB^2(\rho_+ + \rho_-) +\frac{c_2}{2} :{\bf F}^2: \right].
\end{equation}

\begin{figure}
\includegraphics[width=8.5cm]{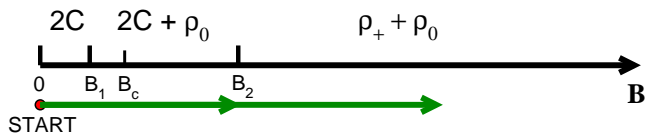}
\caption{Ground state phase diagram of an antiferromagnetic condensate for magnetization $M=N/2$.
We increase $B$ linearly during the time $\tauQ$ to drive the system through a phase transition
into a phase separated state. 
}
\label{scenario}
\end{figure}

Except for the special cases $M=0,\pm N$, the ground state phase diagram, shown in Fig.~\ref{scenario}, 
contains three phases divided by two critical points at
\begin{equation}
B_1=B_0 \frac{M}{\sqrt{2} N},~~
B_2=B_0 \frac{1}{\sqrt{2}},
\end{equation} 
where $B_0=\sqrt{c_2 \rho/A}$ and $\rho$ is the total density. The ground state can be 
$(i)$ antiferromagnetic (2C) with $\psi=(\psi_+,0,\psi_-)$ for $B < B_1$,
$(ii)$ phase separated into two domains of the 2C and $\psi=(0,\psi_0,0)$ type ($\rho_0$) for $B\in(B_1, B_2)$, or
$(iii)$ phase separated into two domains of the $\rho_0$ and $\psi=(\psi_+,0,0)$ type ($\rho_+$) for $B>B_2$~\cite{Matuszewski_PS}.
What is more, the antiferromagnetic 2C state remains dynamically stable, i.e., it remains a local energy minimum up to a critical 
field $B_c>B_1$. Consequently, the system driven adiabatically from the 2C phase, across the phase boundary $B_1$, and into the 
separated phase remains in the initial 2C state up to $B_c>B_1$ when the 2C state becomes dynamically unstable towards
the phase separation.

\begin{figure}
\includegraphics[width=8.5cm]{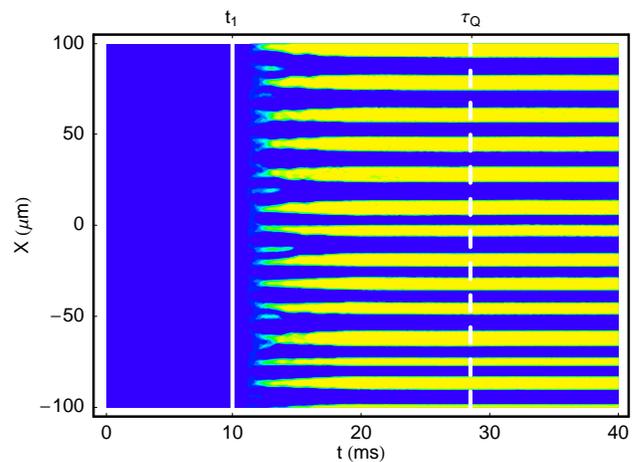}
\caption{
Spin domain formation dynamics in a ring-shaped 1D geometry with ring length $L=200\,\mu$m and $\omega_\perp=2\pi \times 1000\,$Hz,
for $N=2\times 10^{7}$ atoms. The density of the $m_f=0$ component $|\psi_0|^2$ is shown. The quench time is $\tauQ=28.6\,$ms.
}
\label{snapshot}
\end{figure}

For simplicity, we consider a system in the ring-shaped quasi-1D geometry with periodic boundary conditions at $\pm L/2$ and $V(x)=0$.
The magnetic field is initially switched off, and the atoms are prepared in the antiferromagnetic (2C) ground state with magnetization 
fixed to $M=N/2$ (without loss of generality). To investigate KZM we increase $B$ linearly as
\begin{equation}
B(t)~=~B_0~\frac{t}{\tauQ},
\label{tauQ}
\end{equation}
to drive the system through one or two phase transitions into a phase separated state. 
At $B=B_c$, the system is expected to undergo a spatial symmetry breaking phase transition due to
the phase separation into two components. According to the Kibble-\.Zurek theory, 
due to the finite quench time the phase transition has a nonequilibrium
character, and the system ends up in a state with multiple spin domains. At $B=B_2$, on the other hand,
there is no symmetry breaking and the spin-domain landscape remains intact.

The concept of KZM relies on the fact that the system does not follow the ground state exactly in the 
vicinity of the critical point due to the divergence of the relaxation time. 
The dynamics of the system cease to be adiabatic at 
$t\simeq-\hat t$
(here we choose $t=0$ in the first critical point),
when the relaxation time becomes comparable to the inverse quench rate
\begin{equation} \label{tau_freeze}
\hat{\tau}_{\rm rel} \approx |\hat{\varepsilon} / \hat{\dot{\varepsilon}}|,
\end{equation}
where 
$\varepsilon(t)=B - B_c\sim t/\tauQ$
is the distance of 
the system from the critical point.
At this moment, the fluctuations approximately freeze, until the relaxation time
becomes short enough again. After crossing the critical point, distant parts of the system choose 
to break the 
symmetry in different ways, which
leads to the appearance of multiple defects in the form of 
domain walls between domains of 
2C and $\rho_0$ phases. The average number of
domains is related to the correlation length 
$\hat\xi$ 
at the freeze out time 
$\hat{t} \sim \tauQ^{z\nu / (1+ z\nu)}$~\cite{Zurek,Nonequilibrium}
\begin{equation} \label{KZ_scaling}
N_{\rm d} = L / \hat{\xi} \sim \tauQ^{-\nu/(1+z\nu)},
\end{equation}
where $z$ and $\nu$ are the critical exponents determined by the scaling of the relaxation time 
$\tau_{\rm rel} \sim |\varepsilon|^{-z\nu}$ and excitation spectrum $\omega \sim |k|^z$, with $z=1$ in the superfluid.

We test the above prediction in numerical simulations within the truncated Wigner approximation,
with a large number of atoms $N=20\times 10^{6}$ in order to minimize merging of domains thanks to the strong repulsive interaction.
Other parameters are close to that of previous experiments in $^{23}$Na~\cite{Miesner}.
The stochastic equations in the limit of large atom number are equivalent to the time-dependent Gross-Pitaevskii equations
\bea \label{Wigner}
i\hbar\frac{\partial \psi_0}{\partial t}  &=& \left(-\frac{\hbar^2 \nabla^2}{2m} + c_0\rho\right)\psi_0 + \nonumber\\
      &&+c_2\left[(\rho_++\rho_-)\psi_0+2\psi_0^*\psi_+\psi_-\right],\nonumber\\
i\hbar\frac{\partial \psi_+}{\partial t}   &=& \left(-\frac{\hbar^2 \nabla^2}{2m} +c_0\rho+AB^2\right)\psi_+ + \nonumber\\
      &&+c_2\left[(\rho_+-\rho_-)\psi_++\rho_0\psi_++\psi_-^*\psi_0^2\right],\\
i\hbar\frac{\partial \psi_-}{\partial t}   &=& \left(-\frac{\hbar^2 \nabla^2}{2m} +c_0\rho+AB^2\right)\psi_- + \nonumber\\
      &&+c_2\left[(\rho_--\rho_+)\psi_-+\rho_0\psi_-+\psi_+^*\psi_0^2\right],\nonumber
\eea
while the initial condition includes a Wigner-type noise of $1/2$ particle per quantum mode~\cite{WignerRef}.
An example of a single stochastic run, which can be interpreted as a single experimental realization, is shown in Fig.~\ref{snapshot}.
We can clearly see the process of domain formation after the first phase transition at $t_1$.
However,  there is always some number of spin fluctuations that disappear instead of evolving into full domains.
The above dynamics have a striking effect on the number of defects that are created in the system. 
The number of defects and the corresponding scaling law are significantly altered.
In the following, we describe in detail the complete dynamical scenario, going beyond the standard
KZM, and reveal that the post-selection of spin domains is due to the additional conservation law, i.e.~the conservation of magnetization $M$.

\section{Dynamical stability of the initial uniform 2C phase}

We investigate the stability of the uniform 2C state by studying the spectrum of its Bogoliubov excitations~\cite{Leggett_RMP}.
The stationary Gross-Pitaevskii equations derived from the free energy $F=H-\mu N - \gamma M$ simplify to
\bea \label{GPE}
0 &=& \left(c_0\rho-\mu\right)\psi_0 + \nonumber
      c_2\left[(\rho_++\rho_-)\psi_0+2\psi_0^*\psi_+\psi_-\right],\\
0 &=& \left(c_0\rho+AB^2+\gamma-\mu\right)\psi_+ + \nonumber\\
      &&+c_2\left[(\rho_+-\rho_-)\psi_++\rho_0\psi_++\psi_-^*\psi_0^2\right],\\
0 &=& \left(c_0\rho+AB^2-\gamma-\mu\right)\psi_- + \nonumber\\
      &&+c_2\left[(\rho_--\rho_+)\psi_-+\rho_0\psi_-+\psi_+^*\psi_0^2\right],\nonumber
\eea
where $\mu$ is the chemical potential and $\gamma$ is a Zeeman-like Lagrange multiplier to enforce the desired magnetization.
In the 2C state we have $\psi_0=0$ and we can assume, without loss of generality, that both $\psi_+$ and $\psi_-$ are
real and positive. To enforce the desired density and magnetization we set the chemical potential $\mu=c_0\rho+AB^2$ 
and $\gamma=-c_2\rho m_0$. 
Here $m_0$ is relative magnetization 
\be 
m_0~=~\frac{M}{N}.
\ee 
We assume the incompressible
regime $c_0\gg c_2$. After linearization of the time-dependent Gross-Pitaevskii equation in small fluctuations 
$\delta\psi_j(t,x)$ around this uniform background we find that the fluctuations $\delta\psi_0$ decouple from $\delta\psi_\pm$.

The fluctuations $\delta\psi_\pm$ further decouple into the phonon and magnon branch,
\bea
&&
\left(
\begin{array}{c}
\delta\psi_+ \\
\delta\psi_-
\end{array}
\right)= \\
&&
\left(
\begin{array}{c}
\sqrt{\rho_+} \\
\sqrt{\rho_-}
\end{array}
\right)
\int\frac{dk}{\sqrt{2\pi\rho}}
~\left(
b_k^{(p)} u_k^{(p)} e^{ikx}+
b_k^{(p)*} v_k^{(p)*} e^{-ikx}
\right)
~+~\nonumber\\
&&
\left(
\begin{array}{c}
\sqrt{\rho_-} \\
-\sqrt{\rho_+}
\end{array}
\right)
\int\frac{dk}{\sqrt{2\pi\rho}}
\left(
b_k^{(m)} u_k^{(m)} e^{ikx}+
b_k^{(m)*} v_k^{(m)*} e^{-ikx}
\right),\nonumber
\eea
with quasiparticle energies
\bea 
\epsilon_k^{(p)}&=&c_2\rho\sqrt{\xi_{\rm s}^2k^2\left[2(c_0/c_2)+\xi_{\rm s}^2k^2\right]},~~ \nonumber\\
\epsilon_k^{(m)}&=&c_2\rho\sqrt{\xi_{\rm s}^2k^2(8 n_+ n_-+\xi_s^2k^2)},
\eea
respectively, and normalized modes that satisfy
\bea 
u_k^{(p)}\pm v_k^{(p)}&=&\left(\frac{\xi_{\rm s}^2k^2}{2(c_0/c_2)+\xi_{\rm s}^2k^2}\right)^{\pm1/4}~,\nonumber\\
u_k^{(m)}\pm v_k^{(m)}&=&\left(\frac{\xi_{\rm s}^2k^2}{2n_+ n_- +\xi_{\rm s}^2k^2}\right)^{\pm1/4}~,
\eea
where $n_\pm=\rho_\pm/\rho$. Here we use the spin healing length $\xi_{\rm s}=\hbar/\sqrt{2mc_2\rho}$.
The magnon and phonon quasiparticle energies are real and non-negative for any magnetic field $B$. 
There is no instability with respect to the $\delta\psi_\pm$ fluctuations.

The small quadrupole mode fluctuations~\cite{Ho_PRL_1998}
\be 
\delta\psi_0~=~
\int\frac{dk}{\sqrt{2\pi}}
\left(
b_k^{(0)} u_k^{(0)} e^{ikx} + b_k^{(0)*} v_k^{(0)*} e^{-ikx}
\right)
\ee
determine the universality in dynamics of the system. Their quasiparticle energies are
\be 
\epsilon_k^{(0)}~=~c_2\rho\sqrt{[\xi_{\rm s}^2k^2+(1-b^2)]^2-(1-b_c^2)^2}
\label{epsilonk0}
\ee
and the normalized eigenmodes are 
\be 
u_k^{(0)}\pm v_k^{(0)}~=~
\left(
\frac{(b_c^2-b^2)+\xi_{\rm s}^2k^2}{2(1-b_c^2)+(b_c^2-b^2)+\xi_{\rm s}^2k^2}
\right)^{\pm1/4}
\ee
Here we use a rescaled dimensionless magnetic field
\be 
b~=~\frac{B}{B_0}~.
\ee 
The quasiparticle spectrum (\ref{epsilonk0}) is real and positive (finite gap) as long as $b<b_c=\frac{B_c}{B}$, where
\be 
b_c^2~=~1-\sqrt{1-m_0^2}~.
\ee 
At the critical field $b_c$ the gap closes for $k=0$, and above $b_c$ quasiparticle energies for small $k$
become imaginary and the uniform 2C phase develops dynamical instability against the long-wavelength 
$\delta\psi_0$ fluctuations. We note that $b_c>b_1$, where $b_1=B_1/B_0=m_0/\sqrt{2}$, hence there
is a region of bistability of the pure 2C state and the phase separated 2C+$\rho_0$ ground state. The difference 
between $b_1$ and $b_c$, which is the size of this parameter region, is small for weak magnetization $m_0\ll1$. 

In the linear quench (\ref{tauQ}) the state of the system remains in the uniform 2C state as long as that 
state is dynamically stable, i.e., up to the critical field $b_c$. Above $b_c$ it becomes dynamically
unstable against decay towards the mixed $\rho_0$+2C phase. This phase breaks the spatial symmetry,
and thus formation of finite number of defects (domain walls) can be expected.
The process begins by a quasi-exponential
growth of the $\delta\psi_0$ fluctuations on the Kibble-\.Zurek time scale~\cite{Nonequilibrium}
\be 
\hat t ~\sim~\tauQ^{z\nu / (1+ z\nu)} \sim \tauQ^{1/3}~.
\label{hatt}
\ee 
where $z$ and $\nu$ are the critical exponents of the phase transition, $z=1$ and $\nu=1/2$.
By the time $\hat t$ after $b_c$ the $\delta\psi_0$ fluctuations become large, the linearization of the Gross-Pitaevskii
equation breaks down, and the exponential growth of the fluctuations is halted by nonlinearities.
A more accurate estimate for $\hat t$ requires solving the linearized problem.

The dynamics of domain formation is illustrated in Fig.~\ref{snapshot} In the following, we will describe in detail
the physics behind this process, including the post-selection of domains due to the conservation of the magnetization $M$. 
We will demonstrate that the dynamics in the vicinity of the critical point, as well as the long time dynamics, display universal
behavior, but on different spatial scales, determined by two independent scaling laws.

\section{Quasi-exponential growth of the instabilities}

Small fluctuation $\delta\psi_0(t,x)$ around the uniform 2C background satisfies a linearized equation
\be 
i\partial_u\delta\psi_0~=~
-\partial_s^2 \delta\psi_0 +
(1-\epsilon)\delta\psi_0+\delta\psi_0^*.
\ee
Here we use a dimensionless time-like variable $u=tc_2\rho(1-b_c^2)/\hbar$, 
a dimensionless length-like coordinate $s=x~\sqrt{2mc_2\rho(1-b_c^2)/\hbar^2}$, 
and we measure distance from the critical point $b_c$ by a dimensionless parameter $\epsilon=1-(1-b^2)/(1-b_c^2)$.
With the Bogoliubov expansion
\be 
\delta\psi_0(t,s)~=~\int_{-\infty}^\infty\frac{dk}{\sqrt{2\pi}}\left[b_k u_k(t)e^{iks}+b_{k}^*v_{k}^*(t)e^{-iks}\right]
\ee
the linearized equation separates into
\be 
i\partial_u
\left(
\begin{array}{c}
u_k\\
v_k
\end{array}
\right)~=~
\left(
\begin{array}{cc}
1-\epsilon_k & 1 \\
-1 & -1+\epsilon_k
\end{array}
\right)
\left(
\begin{array}{c}
u_k\\
v_k
\end{array}
\right),
\label{bdg}
\ee
where $\epsilon_k=\epsilon-k^2$. 

For each $k$ we need to consider only a solution in the neighborhood of $\epsilon_k=0$. This is the point 
where the mode $k$ becomes dynamically unstable. For $k=0$ the instability is at $\epsilon=0$, that is
$b=b_c$. Other modes become unstable later for $b>b_c$. For a small negative $\epsilon_k$ the positive 
frequency eigenmode of the operator on the right hand side of Eq. (\ref{bdg}) is
\be 
\left(
\begin{array}{c}
u_k\\
v_k
\end{array}
\right)~=~
\frac{1}{\sqrt{2\sqrt{-2\epsilon_k}}}
\left(
\begin{array}{c}
1\\
-1
\end{array}
\right)~.
\label{uvgs}
\ee
This state corresponds to the ground state without the quasiparticle of momentum $k$. This state is
the asymptote of the solution long time before crossing the point of dynamical instability at 
$\epsilon_k=0$.

We consider a linear quench $b(t)=\frac{t}{\tauQ}$ in Eq. (\ref{tauQ}) that can be translated to a nonlinear $\epsilon(u)$.
Since we are interested in $\epsilon\approx0$ we can linearize $\epsilon(u)\approx\frac{u}{u_Q}$ 
with $u_Q=\tauQ\frac{c_2\rho(1-b_c^2)^2}{2\hbar b_c}$ for small time $u$ measured with respect to $\epsilon=0$. 
With this linearized $\epsilon(u)$ Eq. (\ref{bdg}) implies two equations
\be
\partial_z^2 u_k = \left(2z+\frac{i}{u_Q^{1/3}}\right) u_k~,~
\partial_z^2 v_k = \left(2z-\frac{i}{u_Q^{1/3}}\right) v_k~.
\ee
Here $z$ is a time-like variable defined by $\epsilon_k=\frac{u}{u_Q}-k^2\equiv \frac{z}{u_Q^{2/3}}$. 
It measures time with respect to the point $\epsilon_k=0$ where the mode $k$ crosses the point of instability.
The solution is a combination of Airy functions
\bea 
u_k(z) &=& iC~{\rm Ai}(z_+)+C~{\rm Bi}(z_+),\nonumber\\
v_k(z) &=& -iC~{\rm Ai}(z_-)-C~{\rm Bi}(z_-),\nonumber
\eea
with a complex constant $C$ and $z_\pm=2^{1/3}z\pm\frac{i}{2^{2/3}u_Q^{1/3}}$. The modulus of the constant is fixed 
by the condition that the asymptote of the solution for a large negative $z<0$ should equal the state
(\ref{uvgs}) up to an arbitrary phase factor:
\be 
|C|^2 ~=~ \frac{\pi u_Q^{1/3}}{2^{4/3}}
\ee 
Here we used the asymptotic forms of the Airy functions in case of large negative argument.

Once $C$ is fixed, we can work out the asymptote for large positive $z$:
\be 
u_k~\approx~-v_k~\approx~C~\frac{e^{(2z)^{3/2}/3}}{\sqrt{\pi}(2^{1/3}z)^{1/4}}~.
\label{asymp}
\ee
In the truncated Wigner approximation the strength of the initial Gaussian Wigner noise $b_k$ is encoded 
in the correlator $\overline{b_k^*b_p}=\delta(k-p)$. With this noise we obtain average fluctuation density
\be 
\overline{|\delta\psi_0(u,x)|^2}~=~
\sqrt{\frac{mc_2\rho(1-b_c^2)}{6\pi\hbar^2}}~
\frac{e^\alpha}{\sqrt{\alpha}}
\ee
with the quasi-exponential time dependence through $\alpha=2^{5/2}(u^3/u_Q)^{1/2}/3$. 
The fluctuations become large at 
\be 
\hat u\simeq u_Q^{1/3}
\ee
corresponding to the time $\hat t\sim\tauQ^{1/3}$ after crossing the point of dynamical instability $b_c$.

The solution $\hat u\simeq u_Q^{1/3}$ together with the definition of $z$ and the asymptotes (\ref{asymp}) yields 
a power spectrum of the fluctuations proportional to 
\be 
|u_k|^2\approx |v_k|^2 \simeq \exp\frac13\left[2(1-u_Q^{2/3}k^2)\right]^{3/2}
\ee
The spectrum is cut off by the Kibble-Zurek length 
\be 
\hat\xi~\sim~\tauQ^{1/3}~.
\label{hatxi}
\ee
We can conclude that the quasi-exponential growth of the instability halts at the time $\hat t\sim\tauQ^{1/3}$ after the
occurrence of the first dynamical instability at $b_c$. At $\hat t$ the halted fluctuations $\delta\psi_0$ have 
a characteristic Kibble-Zurek length scale $\hat\xi$. The halted fluctuations are potential seeds for $\rho_0$ 
domains in the non-uniform 2C$+\rho_0$ phase. 

\begin{figure}
\includegraphics[width=8.5cm]{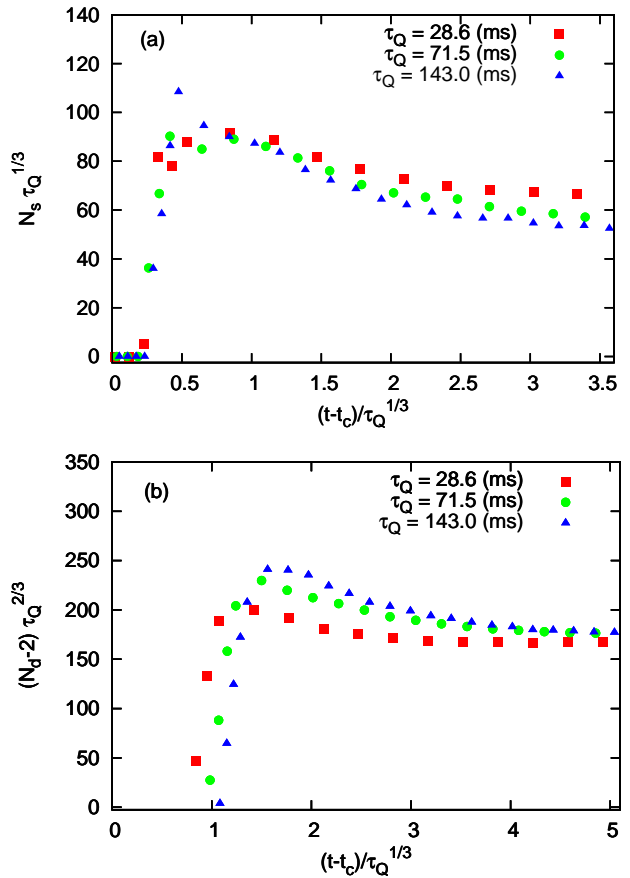}
\caption{Rescaled number of domain seeds $N_{\rm s}$ and number of domains $N_{\rm d}$~\cite{counting_fluctuations} versus rescaled 
time after $t_c$. Here $t_c$ corresponds to $B=B_c$. The figures demonstrate double universal dynamics during the 
formation of domain seeds (top), and at long time (bottom), 
for experiments with different $\tauQ$. 
We account the deviations of rescaled $N_s$ in the top figure to the technical difficulty of determining the exact number of seeds before they are fully formed.
The number of domains is decreased by two to account 
for ground-state phase separation into two domains. Averaged over 100 realizations. Parameters are the same as in Fig.~\ref{snapshot}.
}
\label{universality}
\end{figure}
We recover the analytically predicted temporal and spatial scalings in numerical simulations of domain formation
close to the critical point. In Fig.~\ref{universality}(a), we show the number of small spin domain seeds~\cite{counting_fluctuations}
in function of an appropriately rescaled time distance from the point of instability $t_c$. 
When the number of domain seeds is rescaled taking into account the prediction~(\ref{hatxi}), we can see the universal time dependence
for three different values of $\tauQ$ in the first phase of domain formation. However, at later times
we see a clear departure from the $\tauQ^{1/3}$ scaling law, which is replaced by the $\tauQ^{2/3}$ scaling, as shown
in Fig.~\ref{universality}(b). To explain the occurrence of the second scaling, we need to consider the dynamics beyond the linear 
regime, and consider the stability of the phase separated state itself.

\section{ Stability of the 2C+$\rho_0$ phase}

In the non-uniform 2C+$\rho_0$ there is a phase separation into stationary domains of the 2C phase and the $\rho_0$ phase, see Fig.~\ref{snapshot}.
Sufficiently deep inside each domain $\psi_j$ are independent of $x$, and the stationary Gross-Pitaevskii equations are identical as in the 
uniform case, see Eqs.~(\ref{GPE}).
On one hand, inside a $\rho_0$-domain we have $\psi_+=\psi_-=0$ and the equations reduce to
\be 
\mu=c_0\rho_0~.
\label{mu0}
\ee
On the other hand, inside a 2C-domain we have $\psi_0=0$ and the equations become two conditions
\be 
\mu=c_0(\rho_++\rho_-)+AB^2~,~~\gamma=-c_2(\rho_++\rho_-)m_{\rm 2C}~.
\label{mupm}
\ee
Here $m_{\rm 2C}=(\rho_+-\rho_-)/(\rho_++\rho_-)$ is the relative magnetization in the 2C phase. 
Equations~(\ref{mu0}), (\ref{mupm}) describe chemical equilibrium between the coexisting phases.
Moreover, the energy density inside a $\rho_0$-domain must be 
the same as the energy density inside a 2C-domain,
\bea 
&&
\frac12c_0\rho_0^2-\mu\rho_0~=~\frac12c_2m^2_{\rm 2C}+\gamma m_{\rm 2C}+ \label{pressure}\\
&&
\frac12c_0(\rho_++\rho_-)^2-\mu(\rho_++\rho_-)+AB^2(\rho_++\rho_-),\nonumber
\eea
for the pressure between the different phases to vanish. Finally, the fraction $x_0$ of the system occupied by the phase $\rho_0$ 
must satisfy two conditions:
\bea
&&\rho_0x_0+(\rho_++\rho_-)(1-x_0)=\rho,\\
&&m_{\rm 2C}(1-x_0)=m_0~,
\label{x0}
\eea
for the average density and magnetization on the left hand sides to be $\rho$ and $m_0$ respectively. 
The set of equations (\ref{mu0})-(\ref{x0}) defines the equilibrium conditions 
between the coexisting phases.

Equations (\ref{mu0})-(\ref{pressure}) can be solved with respect to densities:
\be 
\rho_0=\frac{AB^2}{2c_0}+\frac{c_2\rho^2m^2_{\rm 2C}}{2AB^2},~~
\rho_++\rho_-=
-\frac{AB^2}{2c_0}+\frac{c_2\rho^2m^2_{\rm 2C}}{2AB^2},
\ee
From now on we assume for the sake of clarity $c_0\gg c_2$. In this regime the density is the same in both phases and equal to the
initial density $\rho$:
\be 
\rho_0~=~
\rho_++\rho_-~=~
\frac{c_2\rho^2m^2_{\rm 2C}}{2AB^2}~=~
\rho~.
\label{rhorho}
\ee
Even though the density is incompressible, there is still non-trivial spin physics. 
The equilibrium is described by two equations:
\bea 
b^2 &=& \frac{m^2_{\rm 2C}}{2}~,\label{BB0}\\
m_0 &=& (1-x_0)m_{\rm 2C}~.\label{x0m}
\eea
The first of them follows from the last equality in (\ref{rhorho}) and the definition of $B_0=\sqrt{c_2\rho/A}$. 
Since the magnetization $m_{\rm 2C}\in[m_0,1]$, the first condition can be met for $B\in[B_1,B_2]$.

Furthermore, the $2C$ domains are dynamically stable. The Bogoliubov dispersion relation for small $\delta\psi_0$ fluctuations
around the uniform 2C background inside a 2C domain is
\be 
\epsilon_k^{(0)}=c_2\rho
\sqrt{
\left(\xi_{\rm s}^2k^2+(1-b^2)\right)^2-(1-m^2_{\rm 2C})
},
\label{omegak2C}
\ee
compare with the corresponding dispersion (\ref{epsilonk0}) in the initial uniform 2C phase. The stability condition is
\be 
b^2~<~1-\sqrt{1-\frac{m^2_{\rm 2C}}{\rho^2}}.
\label{stab2C}
\ee
It is satisfied given the equilibrium condition $b^2=\frac{m^2_{\rm 2C}}{2}$ in Eq. (\ref{BB0}).
It would not be worth mentioning here, if it were not subject to a reinterpretation in the following argument, 
where we reuse the stability condition (\ref{stab2C}) in a non-equilibrium situation.

Finally, expanding the dispersion relation (\ref{omegak2C}) in powers of small $k$ we can find the healing length
\be 
\xi_{\rho_0+\rm 2C}~=~\xi_{\rm s}~ \frac{\sqrt{2(1-b^2)}}{b^2}~
\label{xi}
\ee
in the $\rho_0$+2C ground phase. This healing length sets the width of a domain wall between the 2C and $\rho_0$ domains.
More precisely, the healing length tells us how deeply the density $\rho_0$ penetrates into the 2C-phase.
Thus $\xi_{\rho_0+\rm 2C}$ is the minimal size of a stable $\rho_0$-domain. This width is finite for any value of magnetic
field in the 2C+$\rho_0$ phase. This picture is completed by the characteristic timescale
\be 
\tau_{\rho_0+\rm 2C}~=~\frac{\hbar}{c_2\rho b^2}~
\ee 
that can be also obtained from the dispersion relation. Again, this time-scale is finite everywhere in the 2C+$\rho_0$ phase.

\section{Domain post-selection dynamics}

At this point we have most ingredients to outline the scenario explaining the unexpected $2/3$ scaling
instead of the standard Kibble-Zurek exponent $1/3$. The linear quench goes through the following stages.

\begin{itemize}

\item The initial uniform state 2C remains dynamically stable from $b=0$ until $b=b_c$.

\item The linearized fluctuations $\delta\psi_0$ around the initial uniform 2C state blow up exponentially near
      the time $\hat t\sim\tauQ^{1/3}$ after crossing $b_c$. The time $\hat t$ corresponds to the magnetic field $\hat b$
      that satisfies $\hat b-b_c\sim\tauQ^{-2/3}$. 
      
\item The explosion of the fluctuations is halted by nonlinearities near $\hat b-b_c\sim\tauQ^{-2/3}$.     
      By this time the density $\rho_0$ still has relatively small amplitude: $\rho_0\ll\rho$.
      There are $\rho_0$-domain seeds whose size is set by the KZ correlation length $\hat\xi\sim\tauQ^{1/3}$,
      and their density scales like $\hat\xi^{-1}\sim\tauQ^{-1/3}$. So far everything goes like in the standard KZ mechanism,
      but now the nonlinear bubble formation steps in.
  
\item For large enough $\tauQ$, we have 
      both $\xi_{\rho_0+\rm 2C}\ll\hat\xi$ and $\tau_{\rho_0+\rm 2C}\ll\hat t$. The last condition implies that $\hat b-b_c$ is the longest 
      ``time-scale'' in the process and thus the nonlinear bubble formation after $\hat b$ can be argued to actually happen near $\hat b$.
      Thanks to the conserved magnetization, only some of the $\rho_0$ seeds will develop into full $\rho_0$-bubbles with $\rho_0=\rho$. 
      As a bubble of $\rho_0$ develops, the magnetization in its surrounding 2C phase is increasing until it reaches a threshold value 
      $\hat m$ when the 2C phase becomes stable again. The stability threshold $\hat m$ follows from a variant of the stability condition (\ref{stab2C}):
      \be 
      \hat b^2~=~1-\sqrt{1-\hat m^2}~.
      \label{hatm}
      \ee
      Once the 2C phase regains its stability the development of new $\rho_0$ bubbles is halted, and the 2C magnetization saturates at $\hat m$.
      The conservation law for the magnetization now reads
      \be 
      m_0~=\hat m~(1-\hat x_0)~.
      \label{hatm2}
      \ee
      Here $\hat x_0$ is the fraction of the length of the system occupied by $\rho_0$ domains.      
      For large enough $\tauQ$ we have $\hat b-b_c\ll b_c$, $\hat m-m_i\ll m_i$, and $\hat x_0\ll1$. Starting from the expression $\hat b^2-b_c^2$ 
      we obtain a relation
      \be 
      \frac{\hat b-b_c}{b_c}~\approx~\left(\frac{\hat m}{m_0}-1\right)\frac{b_1^2}{b_c^2(1-b_c^2)}~.
      \ee
      Using this relation in (\ref{hatm}) we obtain
      \be 
      \hat x_0 ~\sim~\tauQ^{-2/3}~.
      \ee

\begin{figure}
\includegraphics[width=8.5cm]{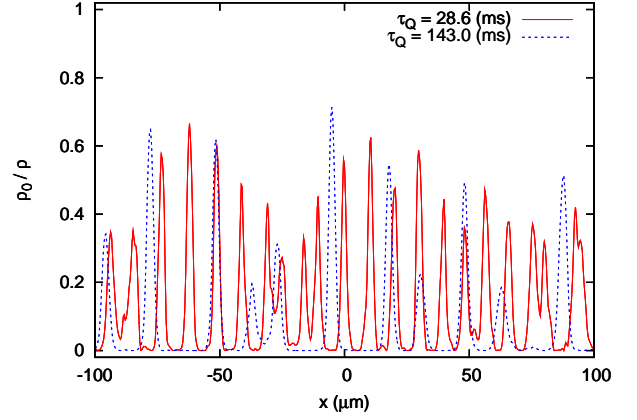}
\caption{Examples of bubble density profiles for two values of $\tauQ$. 
While the average distance between the bubbles is very different in
the two cases, the size of a single bubble remains approximately the same. Parameters as in Fig.~\ref{snapshot}.
}
\label{bubbles}
\end{figure}

\begin{figure}
\includegraphics[width=8.5cm]{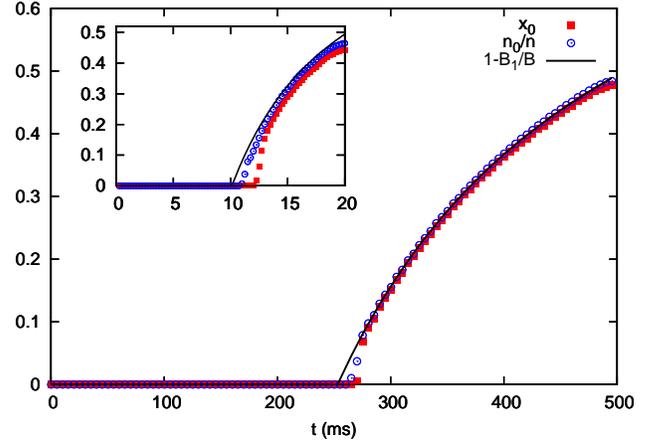}
\caption{Plot of the fraction $x_0$ of the system volume occupied by $\rho_0$ domains, and
the fraction $n_0$ of atoms in the $m=0$ component  
as a function of time in a single experiment. Here $\tauQ=714$ ms for the main plot, and $\tauQ=28.6$ ms for the inset. 
Besides small discrepancies, there is a good agreement with the predicted static value (solid lines). 
The agreement improves with increasing quench time $\tauQ$. Parameters as in Fig.~\ref{snapshot}.
}
\label{ro_zero}
\end{figure}

\item Near $\hat b$ the $\rho_0$-bubbles have the minimal possible size $\simeq\xi_{\rho_0+\rm 2C}$ because, for a given fraction $\hat x_0$ determined
      by the conserved magnetization, such minimal bubbles have the highest possible density 
      \be 
      \frac{N_f}{L}~\simeq~\frac{\hat x_0}{\xi_{\rho_0+\rm 2C}}~\sim~\tauQ^{-2/3}
      \ee 
      and thus their formation requires the system to order over minimal distances. 
      The numerical experiments confirm this result, as shown in Fig.~\ref{bubbles}, which shows the density profiles
      at the time when bubbles are forming from domain seeds, for two different quench times. 
      While the number of bubbles is very different in the two cases, the size of a single bubble is 
      approximately the same. 
      In result, the density of the minimal bubbles scales with the $-2/3$ exponent, in accordance with Fig.~\ref{universality}(b).
         
\item After the formation of minimal bubbles near $\hat b$ the magnetic field keeps growing and the strength of the non-linear Zeeman term in 
      the Hamiltonian is increasing. This increasing term is cooling the system towards it instantaneous ground state, where the $\rho_0$-fraction,
      determined by the conserved magnetization, is 
      \be 
      x_0(b)~=~\frac{b-b_1}{b}. 
      \ee
      The size of the bubbles needs to grow like  
      $
      \frac{x_0(b)}{\hat x_0}~\sim~\frac{b-b_1}{b}
      $ 
      to keep in pace with the increasing $x_0(b)$. This prediction is tested in Fig.~\ref{ro_zero}, where
      we show both the fraction of the length occupied by  $\rho_0$ domains, $x_0$, and the number of atoms
      in the $m=0$ state, $n_0$. These values should coincide if the system separates into perfect
      2C and $\rho_0$ domains. Nevertheless we can see some deviation from the predicted value that is slightly stronger
      for  $x_0$ than $n_0$. This can be explained by the fact that the density in $\rho_0$ and 2C is not exactly the same
      due to the finite ratio $c_0/c_2$, and the fact that 2C domains always contain a small $m=-1$ component. 
      This is especially noticeable at small quench times, see inset in Fig.~\ref{ro_zero}.

\end{itemize}

\begin{figure}
\includegraphics[width=8.5cm]{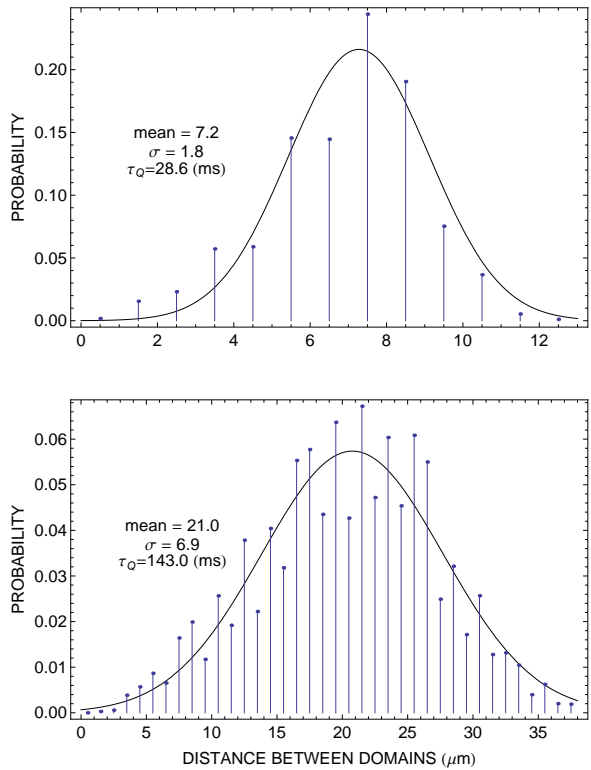}
\caption{Probability of two neighboring domains being separated by a specified distance in micrometers, in 1000 realizations
of the experiments. The solid line is a Gaussian fit. Clearly, there is a regularity in placement of domains. 
Parameters as in Fig.~\ref{snapshot}.
}
\label{histogram}
\end{figure}

The above argument has implications for correlations in the distribution of the minimal $\rho_0$-bubbles along the system. 
When a $\rho_0$-bubble is growing, the conserved magnetization in its neighborhood is increasing making it less favorable 
for another bubble to form there, because the increasing magnetization drives the neighborhood towards the regime of stability of 
the 2C phase. The outcome is effectively the same as if the bubbles repelled: there is anti-bunching in their distribution 
along the system. Crudely speaking, they form something like an imperfect crystal lattice with a preferred ``lattice spacing'' 
distance between the nearest bubbles. This effect is illustrated in Fig. \ref{histogram}. 

\section{Spin domains as quasicondensates}

\begin{figure}
\includegraphics[width=8.5cm]{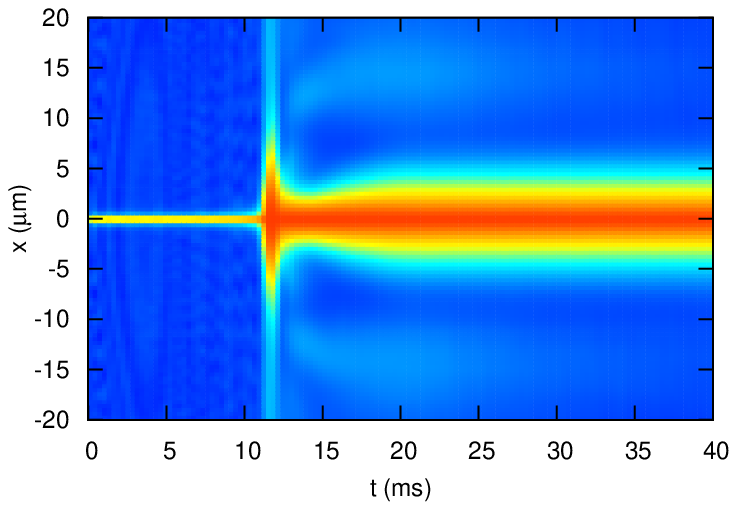}
\includegraphics[width=8.5cm]{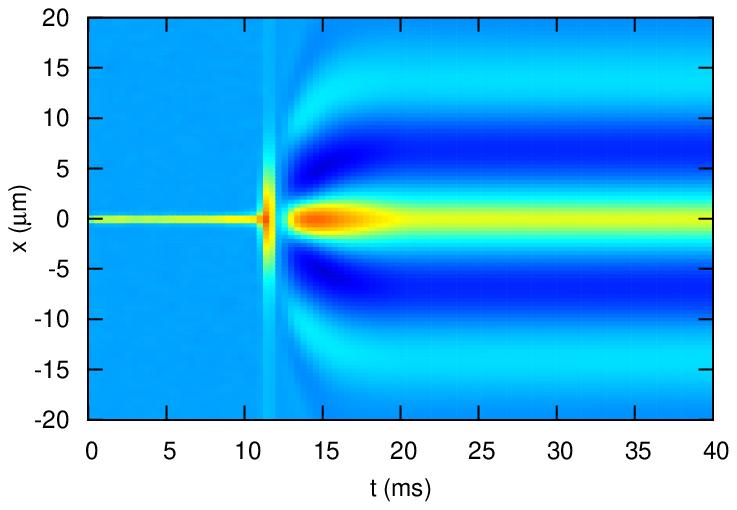}
\caption{Dynamics of correlation functions $g_0^{(1)}(x,0,t)$ (top panel) and  $g_0^{(2)}(x,0,t)$ (bottom panel) 
averaged over $10^4$ realizations for $\tauQ=30$ms, showing the evolution of coherence in the $m_F=0$ spin component.
Parameters as in Fig.~\ref{snapshot}.
}
\label{g1_g2_profiles}
\end{figure}

\begin{figure}[t]
\includegraphics[width=6.8cm, angle=-90]{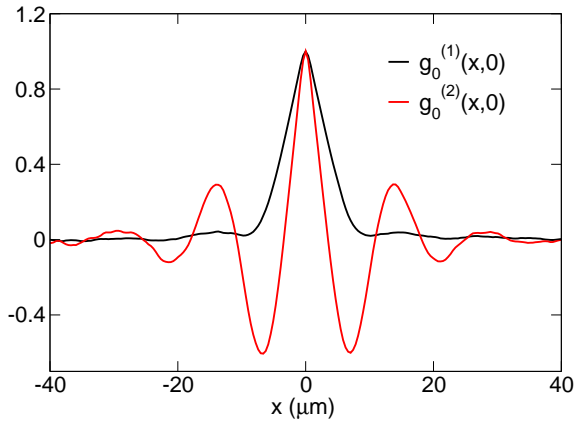}
\caption{Profiles of correlation functions $g_0^{(1)}(x,0,t)$ (black line) and  $g_0^{(2)}(x,0,t)$ 
(red line) at long time (here $t=40$ms), for $\tauQ=30$ms averaged over $10^4$ realizations.
Correlation length can be obtained as the half width of $g_0^{(1)}(x,0,t)$.
The distance between two maxima of 
$g_0^{(2)}(x,0,t)$ corresponds to the average distance between the domains.
}
\label{g1_g2_profiles_long}
\end{figure}

One of the advantages of the truncated Wigner method is the ability to calculate various correlation functions 
in a straightforward way, taking averages over many realizations of the stochastic fields~\cite{WignerRef}.
To investigate the coherence properties of spin domains, we calculated the first- and second-order equal time correlations
\begin{eqnarray}
g_0^{(1)}(x, x',t)&=&\frac{\langle \psi_0^*(x,t) \psi_0(x',t) \rangle}{\sqrt{\langle |\psi_0(x,t)|^2 \rangle \langle |\psi_0(x',t)|^2 \rangle}} \\
g_0^{(2)}(x, x',t)&=&\frac{\langle n_0(x,t) n_0(x',t) \rangle}{\sqrt{\langle n_0(x,t)^2 \rangle \langle n_0(x',t)^2 \rangle}}
\end{eqnarray}
where $n_0(x,t)=|\psi_0(x,t)|^2$. The results for $\tauQ=30$ ms are presented in Figs.~\ref{g1_g2_profiles} and~\ref{g1_g2_profiles_long}.
Before the appearance of domains seeds at $t\approx11$ ms, the state of atoms in the $\psi_0$ component corresponds to a Bogoliubov
vacuum, characterized by a lack of coherence (delta-like correlation function). During the formation of domains, coherence over some
spatial scale is established, and finally the correlation function stabilizes in the shape presented in Fig.~\ref{g1_g2_profiles_long}.
We note that in the absence of domains, the correlation function of $\psi_+$ and $\psi_-$ components displays full coherence,
as the system is in the condensed state. This is due to the finite system size, which allows for condensation
in one dimension~\cite{Bagnato_1DCondensation}.

Closer investigation of the above correlation functions allows us to make some interesting conclusions about the resulting spin domain state.
The shape of $g_0^{(2)}$, with slowly decaying oscillating tails,
is characteristic for systems such as liquids or amorphous solids, which display local anti-correlations of density fluctuations, but 
no density long-range order. This is consistent with the apparent semi-regularity
of domain positions shown in Fig.~\ref{histogram}.
On the other hand, the function $g_0^{(1)}$
does not follow this pattern, as there are almost no oscillations and the spatial decay is much faster.
In fact, the first-order correlation function decays to zero on a length scale corresponding to the distance between neighboring domains.
This indicates that there is no phase coherence between the domains, and consequently, spin domains can be seen as
a set of quasicondensates. This effect is due to the existence the insulating 2C phase between the $\rho_0$ domains, 
which prevents tunneling of $\psi_0$ atoms and phase locking. The same effect can be seen for domains of the 2C phase.

\section{Conclusion}

We described in detail formation of $\rho_0$ domains in a transition from the antiferromagnetic 2C phase to the separated $\rho_0$+2C phase.
As the control parameter (magnetic field) is turned on, it crosses the critical value $B_c$ when the 2C phase becomes dynamically unstable towards the
exponential growth of $\rho_0$ fluctuations. These fluctuations are seeds for the $\rho_0$ domains in the phase separated phase. The very passage across this 
instability is described by the Kibble-Zurek (KZ) theory and, in particular, the density of the $\rho_0$ seeds scales according to the KZ scaling laws. However,
it is impossible for most of the $\rho_0$ seeds to continue growing until they become the fully fledged $\rho_0$ domains, because their density would be
too high to be compatible with the conserved total magnetization. Thus the $\rho_0$ seeds are subject to a quick (non-linear) post-selection that happens on the
same time-scale $\hat t$ as the KZ mechanism and decimates their density just to satisfy the conservation law. The net outcome is a finite density 
of $\rho_0$-bubbles whose density satisfies a scaling law that is different from the KZ scaling.

The initial size of the $\rho_0$ bubbles does not depend on the transition rate, but after their formation the size grows with the magnetic field in such 
a way that the fraction of the system occupied by the $\rho_0$ phase keeps in pace with the same fraction in the ground state of the $\rho_0$+2C phase for 
a given magnetic field. One of the implications of the post-selection mechanism is that the $\rho_0$ bubbles are positioned in a semi-regular fashion like 
in an imperfect crystal lattice. What is more, there are no phase correlations between different bubbles: they are a set of mutually phase-uncorrelated 
condensates of the $m_f=0$ component. The same is true for the train of 2C-domains that separate the $\rho_0$ condensates. 

\acknowledgments

This work was supported by the Polish Ministry of Science and Education grant
IP 2011 034571, the National Science Center grants DEC-2011/01/B/ST3/00512 (J.D.) and DEC-2011/03/D/ST2/01938,
and by the Foundation for Polish Science through the
\textquotedblleft Homing Plus\textquotedblright\ program.

\clearpage


\begin{thebibliography}{99}

\bibitem{RenormalizationGroup} J. Binney, N. Dowrick, A. Fisher, M. Newman, {\it The Theory of Critical
Phenomena: An Introduction to the Renormalization Group} (Oxford University
Press, Oxford, UK, 1992).

\bibitem{NonequilibriumBooks} G. Nicolis and I. Prigogine, {\it Self-organization in nonequilibrium systems:
from dissipative structures to order through fluctuations} (Wiley, 1977);
H. Haken, {\it Synergetics: An Introduction : Nonequilibrium Phase Transitions and Self-Organization in Physics, Chemistry and Biology}  (Springer, Berlin, 1978); H. Hinrichsen, Physica A {\bf 369}, 1 (2006).

\bibitem{Kibble} T. W. B. Kibble, J. Phys. A {\bf 9}, 1387 (1976);
T. W. B. Kibble, Phys. Rep. {\bf 67}, 183 (1980).

\bibitem{Zurek} W. H. \.Zurek, Nature (London) {\bf 317}, 505 (1985); 
W. H. \.Zurek, Acta Phys. Pol. B {\bf 24}, 1301 (1993);
W. H. \.Zurek, Phys. Rep. {\bf 276}, 177 (1996).

\bibitem{Helium_KZ} C. Bauerle, Y.~M.~Bunkov, S.~N.~Fisher, H.~Godfrin, and G.~R.~Pickett, Nature
(London) {\bf 382}, 332 (1996);
V.M.H. Ruutu, V.~B.~Eltsov, A.~J.~Gill, T.~W.~B.~Kibble, M.~Krusius, 
Y.~G.~Makhlin, B.~Pla{\c c}ais, G.~E.~Volovik, and W.~Xu, ibid. {\bf 382}, 334 (1996).

\bibitem{other_KZ} I. Chuang , B.~Yurke, R.~Durrer, and N.~Turok, Science {\bf 251}, 1336 (1991).

\bibitem{superconductors_KZ} A. Maniv, E. Polturak, and G. Koren, Phys. Rev. Lett. {\bf 91}, 197001 (2003); 
R. Monaco, J.~Mygind, M.~Aaroe, R.~J.~Rivers, and V.~P.~Koshelets,  ibid. {\bf 96}, 180604 (2006).

\bibitem{BEC_KZ}  L. E. Sadler, J. M. Higbie, S. R. Leslie, M. Vengalattore, and D. M. Stamper-Kurn, Nature (London) 443, 312 (2006).
                  D. R. Scherer, C.~N.~Weiler, T.~W.~Neely, and B.~P.~Anderson,
                  Phys. Rev. Lett. {\bf 98}, 110402 (2007);
                  R. Carretero-Gonzalez, B.~P.~Anderson, P.~G.~Kevrekidis, 
D.~J.~Frantzeskakis, and C.~N.~Weiler,
                  Phys. Rev. A {\bf 77}, 033625 (2008);
                  C. N. Weiler,  T.~W.~Neely, D.~R.~Scherer, A.~S.~Bradley, M.~J.~Davis, and 
B.~P.~Anderson, Nature (London) {\bf 455}, 948 (2008); 
                  E. Witkowska, P. Deuar, M. Gajda, and K. Rza\.zewski, Phys. Rev. Lett. {\bf 106}, 135301 (2011).
                  D. Chen, M. White, C. Borries, and B. DeMarco,
                  arXiv:1103.4662.

\bibitem{Esslinger} K. Baumann, R. Mottl, F. Brennecke, and T. Esslinger,
                    Phys. Rev. Lett. {\bf 107}, 140402 (2011).      
                    
\bibitem{Ulms} K. Pyka, J. Keller, H. L. Partner, R. Nigmatullin, T. Burgermeister, 
               D.-M. Meier, K. Kuhlmann, A. Retzker, M. B. Plenio, W. H. Zurek, A. del Campo, T. E. Mehlst\"ubler,
               arXiv:1211.7005;
               S. Ulm, J. Rossnagel, G. Jacob, C. Degünther, S.T. Dawkins, U.G. Poschinger, R. Nigmatullin, 
               A. Retzker, M.B. Plenio, F. Schmidt-Kaler, K. Singer, 
               arXiv:1302.5343.    


\bibitem{QPT_Book}
S. Sachdev, Quantum Phase Transitions (Cambridge University Press, Cambridge UK, 2001).

\bibitem{Other_QPT}
J. Dziarmaga, Phys. Rev. Lett. {\bf 95}, 245701 (2005);
W. H. \.Zurek, U. Dorner, and P. Zoller, Phys. Rev. Lett. {\bf 95}, 105701 (2005); 
B. Damski, Phys. Rev. Lett. {\bf 95}, 035701 (2005).

\bibitem{Spinor_FerromagneticKZ} A. Lamacraft, Phys. Rev. Lett. {\bf 98}, 160404 (2007);
B. Damski and W. H. \.Zurek, Phys. Rev. Lett. {\bf 99}, 130402 (2007);
J. Sabbatini, W. H. \.Zurek, and M. J. Davis, Phys. Rev. Lett. {\bf 107}, 230402 (2011);
J. Sabbatini, W. H. \.Zurek, and M. J. Davis, New J. Phys. {\bf 14}, 095030 (2012).

\bibitem{Nonequilibrium}
J. Dziarmaga, Adv. Phys. {\bf 59}, 1063 (2010);
A. Polkovnikov, K. Sengupta, A. Silva, M. Vengalattore, Rev. Mod. Phys. {\bf 83}, 863 (2011).

\bibitem{Nasz_PRL} T. \'Swis\l{}ocki, E. Witkowska, J. Dziarmaga, M. Matuszewski, Phys. Rev. Lett. {\bf 110}, 045303 (2013).

\bibitem{Chang_NP_2005} M. S. Chang, Q. S. Qin, W. X. Zhang, L. You, and M. S. Chapman, Nat. Phys. {\bf 1}, 111 (2005).

\bibitem{Matuszewski_AF} M. Matuszewski, T. J. Alexander, and Y. S. Kivshar,
Phys. Rev. A {\bf 78}, 023632 (2008); 

\bibitem{Matuszewski_PS} M. Matuszewski, T. J. Alexander, and Y. S. Kivshar,
Phys. Rev. A {\bf 80}, 023602 (2009).
M. Matuszewski, Phys. Rev. Lett. \textbf{105}, 020405 (2010);
M. Matuszewski, Phys. Rev. A {\bf 82}, 053630 (2010).

\bibitem{Miesner} 
J. Stenger, S. Inouye, D. M. Stamper-Kurn, H.-J. Miesner, A. P. Chikkatur, and W. Ketterle, 
Nature (London) {\bf 396}, 345 (1998);
H.-J. Miesner, D. M. Stamper-Kurn, J. Stenger, S. Inouye, A. P. Chikkatur, and W. Ketterle,
Phys. Rev. Lett. {\bf 82}, 2228 (1999);

\bibitem{WignerRef} M. J. Steel, M. K. Olsen, L. I. Plimak, P. D. Drummond, S. M. Tan, M. J. Collett, D. F. Walls, and R. Graham, Phys. Rev. A {\bf 58}, 4824 (1998); A. Sinatra, C. Lobo, and Y. Castin, Phys. Rev. Lett. {\bf 87}, 210404 (2001).

\bibitem{Leggett_RMP} A. J. Leggett, Rev. Mod. Phys. {\bf 73}, 307 (2001).

\bibitem{Ho_PRL_1998} T.-L.~Ho, Phys.~Rev.~Lett.~{\bf 81}, 742 (1998); T. Ohmi and K. Machida, J. Phys. Soc. Jpn. {\bf 67}, 1822 (1998).

\bibitem{counting_fluctuations} To determine
the number of domains $N_{\rm d}$  or domain seeds $N_{\rm s}$ we count the number of zero crossings of 
the function $f(x)=n_0(x) - \alpha N/L$, where $\alpha=0.5$ for domains $\alpha=0.03$ for domain seeds,
at the time instant when $N_{\rm s}$ or  $N_{\rm d}$ is the largest. We checked that this method is accurate and weakly dependent 
on the choice of $\alpha$.

\bibitem{Bagnato_1DCondensation} V. Bagnato and D. Kleppner, Phys. Rev. A {\bf 44}, 7439 (1991); 
I. Bouchoule, K. V. Kheruntsyan, and G. V. Shlyapnikov,  Phys. Rev. A {\bf 75}, 031606(R) (2007).



\end{thebibliography}
\end{document}